\documentclass[a4paper,11pt]{article}
\usepackage[utf8]{inputenc}

\usepackage{geometry}
\usepackage{xcolor}
\usepackage{amsmath, array, amssymb, amsfonts,amsthm}
\usepackage{xfrac}
\usepackage[inline]{enumitem}

\usepackage[french, english]{babel}
\usepackage[all]{xy}

\usepackage{txfonts}  

\usepackage{upgreek}

\usepackage{sectsty}
\sectionfont{\centering}
\subsectionfont{\centering}
\subsubsectionfont{\centering}

\usepackage{booktabs}
\usepackage{caption}
\usepackage{dsfont}
\usepackage{mathtools}

\usepackage[hidelinks]{hyperref}

\usepackage{textcomp}

\usepackage{multicol}
\usepackage[square,numbers,sort,compress,semicolon,merge]{natbib}
\let\cite\citep 
\usepackage{hyperref}
\hypersetup{colorlinks=true, urlcolor=blue, citecolor=blue, linktoc=page}

\renewcommand\P{\mathcal{P}}
\newcommand\M{\mathcal{M}}

\newcommand\RR{\mathbb{R}}
\newcommand\CC{\mathbb{C}}
\renewcommand\1{\textbf{1}}

\renewcommand\H{\mathcal{H}}
\newcommand\A{\mathcal{A}}

\newcommand\SU{\mathcal{SU}}
\newcommand\U{\mathcal{U}}
\newcommand\SO{\mathcal{SO}}

\newcommand\J{\mathcal{J}}

\newcommand\vphi{\varphi}

\renewcommand\epsilon{\varepsilon}

\newcommand\rarrow{\rightarrow}

\newcommand*\vect[1]{\begin{pmatrix}#1 \end{pmatrix}}

\newcommand\LieH{\mathfrak{h}}

\newcommand\so{\mathfrak{so}}

\renewcommand\t{\widetilde}

\renewcommand\b{\bar }
\newcommand\w{\wedge}
\renewcommand\d{\partial}

\renewcommand\-{^{-1}}

\renewcommand\1{\mathds{1}}

\DeclareMathOperator{\Tr}{Tr}
\DeclareMathOperator{\vol}{vol}

\begin{document}


\title{Artificial \emph{vs} Substantial Gauge Symmetries: a Criterion and an Application to the Electroweak Model}

\author{J. François${\,}^{a,b}$}
\date{}

\maketitle
\vskip -0.5cm
\noindent
${}^a$ Université de Lorraine, Institut {\'E}lie Cartan de Lorraine, UMR 7502, Vandoeuvre-lès-Nancy, F-54506, France\\
${}^b$ CNRS, Institut {\'E}lie Cartan de Lorraine, UMR 7502, Vandoeuvre-lès-Nancy, F-54506, France

\begin{abstract}
To systematically answer the generalized Kretschmann objection, we propose a mean to make operational a  criterion widely recognized as allowing to decide if the gauge symmetry of a theory is artificial or substantial. Our proposition is based on the dressing field method of gauge symmetry reduction, a new simple tool from mathematical physics. This general scheme allows in particular to straightforwardly argue that the notion of spontaneous symmetry breaking is superfluous to the empirical success of the electroweak theory. Important questions regarding the context of justification of the theory then arise. 
\end{abstract}


\vspace{1.5mm}



\section{Introduction}  

The philosophical analysis of gauge symmetries, long overdue, gained particular interest in the past fifteen years. Several notions deserve scrutiny. One is the gauge principle, according to which imposing a local/gauge symmetry on a free theory turns it into an interacting theory. This was suggestively encapsulated by Yang's aphorism ``Symmetry dictates interaction'' \cite{YangSelPap}, clearly one of the conceptual revolution of the last century. But not long ago philosophers of physics took hold of the celebrated principle and scrutinized it (see e.g \cite{Martin2002,Castellani-Brading2003, Lyre2004}), as did some dissenting physicists much earlier, see e.g \cite{OP1962}. It was quickly found that actually gauge symmetries might not be the sole criterion constraining the space of admissible theories and that others like, renormalizability, may be more fundamental.

Nevertheless, it seems undeniable that gauge symmetries are a powerful heuristic guide to find fruitful and ultimately empirically adequate theories of the fundamental interactions. So, they appear to tell something deep and important about Nature. But this conclusion raises a well known problem, which can be summarized by saying that there is a generalized Kretschmann objection applicable to gauge symmetries. As a reminder, shortly after Einstein delivered his General Theory of Relativity (GR), Erich Kretschmann suggested in 1917 that the principle of general covariance was empty, unable to constrain the space of admissible theories, since any theory could be written as to be generally covariant. There has been a long and lively debate over the validity of Kretschmann's objection and the relevance of general covariance in relativity theory, and much effort to determine if there is a demarcation criterion to distinguish artificial general covariance from substantial general covariance \cite{Pooley2017}. It happens that much of this discussion applies, mutatis mutandis, to gauge symmetries and that a generalized Kretschmann objection \cite{Pitts2008, Pitts2009} can be raised against the gauge principle: Physicists have devised many ways to implement a gauge symmetry in a theory lacking it \cite{Ruegg-Ruiz, BatalinTyutin1991}, so if any theory can be turned into a gauge theory how come  gauge symmetries are regarded as a fundamental insight into the hidden structure of Nature? 

One is led to suspect that there are ``native'' gauge theories endowed with substantial gauge symmetries signaling a genuine physical content, and theories whose gauge symmetries are artificial and devoid of physical meaning. Identifying the physical signature of substantial gauge symmetries, even partially, is necessary to determine a demarcation criterion. Once found such a criterion, it may still be another matter to suggest a systematic tool to make it operational.  

It is by now quite widely acknowledged that substantial gauge symmetries are related to locality/non-locality in field theory \cite{Guay2008, Healey2009, Dougherty2017, Nguyenetal2017}.\footnote{ 
As far as field theory is concerned, locality can be seen as consisting in the following three desiderata.
(1) Relativistic causality: Physical processes (carrying energy/information) propagate within the lightcone structure of spacetime.
(2) Field locality: Physical properties (fields) are defined pointwisely (in the idealized limit of arbitrary small regions of spacetime), and interact pointwisely in regions where they do not vanish. It is closely related to 
(3) Separability: Physical properties (fields configurations) of any region of spacetime are recovered from physical properties of its constitutive subregions. 
Relativistic causality is still non-negotiable, so failure of locality means infringing field locality \cite{Healey2009} or separability \cite{Dougherty2017, Nguyenetal2017}, or both \cite{Guay2008}.}
Precisely, substantial gauge theories exhibit a trade-off between gauge symmetry and locality: Either the theory is expressed with local gauge field variables, or with non-local gauge invariant ones. Theories with artificial gauge symmetries do not have this property, they can be expressed in terms of gauge invariant local field variables. 
On this distinction one can draw the conclusion that since, as far as we know, the physical content of a theory is gauge invariant, a substantial gauge symmetry would then signal the existence of non-local phenomena.

Whether or not this notion exhausts the content of gauge symmetries, we can take it as a first robust demarcation criterion. 
But how can it be implemented on any given gauge theory? Absent a general theorem, it appears that the best strategy available is to try to show that the theory can be written in term of gauge invariant local variables. The failure of all efforts to do so ascertains, at least provisionally, the substantial nature of the gauge symmetry of the theory. Success, on the other hand, definitely shows that the gauge symmetry is artificial.   
The problem at hand is then  to make this asymmetric strategy as systematic and operational as possible.

One of the goal of the present paper is to bring attention to a new simple mathematical tool to deal with gauge symmetries, the dressing field method, that might be suited for just such a task and which applies to a wide class of gauge theories, namely those that can be formalized in the language of Lagrangians (and fiber bundles).
 Section \ref{The dressing field method of gauge symmetry reduction} gives a brief description of this method, which is easy to grasp in its gist, and refers to the published literature for technical details and elaborations. 
  Then, to motivate our proposition for an operational criterion based on the method, we compare the massive Stueckerlberg electromagnetic model and the standard massless $U(1)$-gauge theory. A point that is highlighted by the method is that the nature of a gauge symmetry depends essentially on the field content of the theory. This is illustrated by a discussion of the Aharonov-Bohm effect, and by considering the tetrad gauge theoretic formulation of GR. Also, to further probe the scope of the method, we briefly indicate why the so-called ``clock fields'' appearing in massive gravity and bi-gravity theories are natural instances of dressing fields. 
 
\medskip

 Section \ref{The electroweak theory without spontaneous symmetry breaking} articulates the second main point of the paper,  namely that the orthodox interpretation of the electroweak unification is untenable. It is  indeed still common wisdom among practicing physicists to consider the notion of spontaneous symmetry breaking (SSB) as pivotal to the success of the theory.
The idea is often seen as an important insight into the structure of physical reality. This opinion is   shared widely enough  that virtually all modern textbooks on gauge field theory or quantum field theory (QFT) have a chapter on SSB, and that even prominent physicists and science popularizers convey it to the layman \cite{Carroll2012, Carroll2015} (sometimes going as far as to suggest that it is the most revolutionary discovery of  XXth century theoretical physics \cite{Krauss2017, KraussSA2017}).
 
But since at least fifteen years, philosophers of physics have voiced skepticism. Here we provide supporting evidences: By relying on the dressing field method, which gives a clear conceptual language to elucidate the real content  of the electroweak unification,   we argue that the empirical success of the Glashow-Weinberg-Salam model is entirely independent of the interpretation in terms of SSB.  
It is to be hoped that this conclusion 
will come to be more widely acknowledged in the physics community. Unfortunately,  scientists can be somewhat dismissive of the inquiries of philosophers,\footnote{One recalls the notorious example of   \cite{Hawking2010}, or the controversial interview by Lawrence Krauss in The Atlantic \cite{KraussAtlantic2012} where he opinioned that  ``[...] the worst part of philosophy is the philosophy of science; the only people, as far as I can tell, that read work by philosophers of science are other philosophers of science. It has no impact on physics what so ever, and I doubt that other philosophers read it because it's fairly technical. And so it's really hard to understand what justifies it.''  He later gave a retraction in a column of Scientific American \cite{KraussSA2012}.} 
 an attitude for which little price is usually paid, in the short term at least. 
But in this case, not acknowledging the insights of philosophers of physics would certainly lead to a long-lived misconception at the heart of particle physics to remain uncorrected for still some times, and important ensuing questions regarding the context of justification of the electroweak model to remain unasked, let alone answered. 
This is addressed in the concluding remarks of section \ref{Closing statement, open questions}.
 
Notice that throughout the paper we will use the language of differential forms,\footnote{Here, differential forms should be understood broadly as including twisted forms \cite{Burke1985}, or pseudo-forms \cite{Frankel2011}, describing tensor densities needed e.g for integration on non-orientable manifolds (even if usually we will assume  spacetime to be orientable).} first because it is  common practice in mathematical physics and in differential geometry (within which the dressing field method was developed), and also because it frees the mind of cumbersome index notations. I hope that readers unfamiliar with this language will nevertheless follow most of the argument presented without much trouble.

\section{The Dressing Field Method of Gauge Symmetry Reduction}  
\label{The dressing field method of gauge symmetry reduction}

The dressing field method has been devised as a mean to handle, i.e reduce, gauge symmetries in a way that differs markedly from either gauge fixing or SSB mechanisms, and bears some resemblance to the bundle reduction theorem from the differential geometry of fiber bundles (see \cite{Sharpe}, p.147-149, proposition 2.14). It has been applied mainly to conformal geometry where it allows to recover tractor and twistor calculi (analogues for conformal manifolds of the Ricci and spinorial calculi on Riemannian manifolds) from a gauge reduction of the Cartan conformal geometry \cite{Attard-Francois2016_I, Attard-Francois2016_II}. It also uncovered a new class of gauge fields, called \emph{non-standard} or \emph{twisted} gauge fields, which generalize connections and sections of vector bundles used in Yang-Mills theories to model gauge potentials and matter fields. See \cite{Attard_et_al2017} for a review with technical details and references. 
The method is fully developed within the geometry of fiber bundles, but its basics and application to physical models are easy to state.

\subsection{The basic mathematical setup of the method} 
\label{The basic mathematical setup of the method}

 First, we need to lay the setup for a gauge theory. Consider a gauge theory on a $m$-dimensional spacetime manifold $(\M, g)$, whose gauge group of symmetry
 is $\H:=\{\gamma: \M \rarrow H \}$ with $H$ a Lie group with Lie algebra $\LieH$, and which by definition acts on itself via $\gamma_1^{\gamma_2}=\gamma _2\- \gamma_1\gamma_2$.\footnote{Here $\H$  can be seen either as the pulled-back version of the group of vertical automorphisms of the underlying principal bundle $\P(\M, H)$ over $\M$, or as the group of transition functions between different trivializations of  $\P$. Dealing with active or passive gauge transformations makes no formal difference.} 
 The basic space of fields is $\Phi=\{A, F, \vphi\}$, where $F$ is the field strength (curvature $2$-form $\in \Omega^2(\M, \LieH$)) of the gauge potential $A$ (connection $1$-form $\in \Omega^1(\M, \LieH$)), and $\vphi$ is a matter field pertaining to a representation $(\rho, V)$ of $H$. The gauge group acts on the space of fields, $\Phi \xrightarrow{\H}\Phi^\gamma$, as
\begin{align}
\label{GT}
A^\gamma&=\gamma\-A\gamma +\gamma\-d\gamma, \qquad F^\gamma=\gamma\-F\gamma , \notag\\
\vphi^\gamma&=\rho(\gamma\-) \vphi \qquad \text{and} \qquad D\vphi^\gamma=\rho(\gamma\-)D\vphi, 
\end{align}
where $D:=d + \rho_*(A)$ is the covariant derivative implementing the minimal coupling between the matter field and the gauge potential. 

Now, a physical theory is specified by its Lagrangian $m$-form $L(A, \vphi)$. In the case of a gauge theory, the Lagrangian is required to be $\H$-gauge invariant: $L(A^\gamma, \vphi^\gamma)=L(A, \vphi)$. A prototypical and almost minimal Yang-Mills Lagrangian is
\begin{align}
\label{YM}
L(A, \vphi, \vphi')&=L_{\text{YM}}+ L_{\text{Scalar}} + L_{\text{Dirac}}, \notag\\
		    &=\tfrac{1}{2}\Tr(F \wedge *F) + \langle D\vphi, *D\vphi \rangle -m^2\langle\vphi, *\vphi \rangle+ \langle \vphi',  \upgamma \wedge *D\vphi'\rangle - m'\langle\vphi', *\vphi' \rangle,
\end{align}
where $\vphi$ is a scalar field with mass $m$ and $\vphi'$ is a spinor field with mass $m'$. 
Here $\w$ is the wedge product of differential forms, $*:\Omega^p(\M) \rarrow \Omega^{m-p}(\M)$ is the Hodge star operator, while $\Tr$ and $\langle\ ,  \rangle$ are bilinear forms on $\LieH$ and $V$ respectively. As for $\upgamma=\upgamma_\mu dx^\mu$, it is a one form whose components are Dirac gamma matrices. 
 A mass term for the gauge potential A, $\mu^2 Tr(A \w *A)$, failing to be gauge-invariant by virtue of \eqref{GT}, is  forbidden so that a gauge interaction is a priori long range.
 \medskip
 
 The core idea of the dressing field method consists in the following simple observation. Suppose the structure group $H$ has some normal subgroup $J$,\footnote{The subgroup $J$ is normal if $\forall h \in H$ on has $h^{-1}Jh=J$. Said otherwise $\forall h \in H$ and $\forall j \in J: h^{-1}jh \in J$. The requirement of normality for $J$, while not strictly necessary, insures that $H/J$ is still a group. Nothing conceptually important rests on this technical detail. 
 } so that the gauge group $\H$ correspondingly has a normal subgroup $\J$. 
 Suppose further that from the space of fields $\Phi$ of a gauge theory, one can extract a field $u:\M \rarrow J$ defined by the transformation property under $\gamma \in\J$: $u^\gamma=\gamma\-u$.   Such a field is called a \emph{dressing field}. It allows to perform a change of field variables, $\Phi \rarrow \Phi^u$, by forming the following \emph{composite fields} :
 \begin{align}
 \label{CompFields}
 A^u:&=u\- Au +u\-du, \qquad F^u=u\-Fu,   \notag \\
 \vphi^u:&=\rho(u\-) \vphi \qquad \text{and} \qquad D^u\vphi^u=\rho(u\-)D\vphi, 
 \end{align}
 where $D^u:=d +\rho_*(A^u)$. These fields are $\J$-invariant variables. 
 Notice that  despite a formal similarity with \eqref{GT}, \eqref{CompFields} are \emph{not} gauge transformations. Indeed, by virtue of its defining transformation property, the dressing field is not an element of the gauge group: $u\notin \H$.
 
 Taking advantage of the $\H$-gauge invariance of the Lagrangian, which holds as a strictly formal property, and of the formal similarity between \eqref{GT} and \eqref{CompFields}, we can rewrite the gauge theory in terms of the $\J$-invariant variables:
 \begin{align*}
 L(A, \vphi, \vphi')&=L(A^u, \vphi^u, {\vphi'}^u),  \notag \\
 				  &=\tfrac{1}{2}\Tr(F^u \wedge *F^u) + \langle D^u\vphi^u, *D^u\vphi^u \rangle -m^2\langle\vphi^u, *\vphi^u \rangle+ \langle {\vphi'}^u, \upgamma \wedge *D^u{\vphi'}^u\rangle - m'\langle{\vphi'}^u, *{\vphi'}^u \rangle.
 \end{align*}
This theory is therefore not a $\H$-gauge theory, as the  Lagrangian \eqref{YM}  would have us think, but a $\H/\J$-gauge theory.
Clearly the transformation of the composite fields \eqref{CompFields} under the residual $\H/\J$-gauge symmetry depends on the  behavior of the dressing field under this same symmetry. Interesting cases are described in \cite{Attard_et_al2017}.

Insofar as the genuine physical degrees of freedom (d.o.f) of a gauge theory are given by gauge invariant quantities, the dressing field method helps to exhibit the physical content of a gauge theory.
\bigskip

 A question naturally arises: Is the field redefinition \eqref{CompFields} harmless? 
As already observed above, formally the dressing operation exactly mimics gauge transformations \eqref{GT} which form a symmetry of the theory. So, it is expected that at the classical level the theories expressed with the original variables and the dressed ones are  equivalent: The field equations for the composite fields $\{A^u, \vphi^u, {\vphi'}^u  \}$ obtained from $L(A^u, \vphi^u, {\vphi'}^u)$ are the same as those of the original gauge variables $\{A, \vphi, \vphi'\}$ obtained from $L(A, \vphi, \vphi')$. Both are in a relation of ``covariance''  with respect to (w.r.t) the dressing operation by $u$, in the same way that the field equations of gauge related fields $\{A, \vphi, \vphi' \}$ and  $\{A^\gamma, \vphi^\gamma, {\vphi'}^\gamma \}$  are in a relation of covariance w.r.t the action of the gauge group $\H $.  

For the same reason, at the quantum level the redefinition \eqref{CompFields} should be harmless if the gauge symmetry is preserved in the quantized theory, i.e if it  is free of gauge anomalies, which is the case for any viable candidate as a fundamental QFT \cite{Bertlmann}. One would then submit that if a gauge QFT is anomaly free, it is equivalent to its dressed counterpart.\footnote{This has some relevance regarding our discussion of the electroweak model on section \ref{The electroweak theory without spontaneous symmetry breaking}. Conversely, if a gauge QFT can be ``dressed" with a dressing field for $\J$ 
(a \emph{local} one, see section \ref{Artificial vs substantial gauge symmetry}), then it is trivially free of $\J$-gauge anomalies. Indeed,  once expressed in terms of the composite fields \eqref{CompFields} there remains no $\J$-gauge symmetry to break in the quantization procedure.}

Now, even if one considers a quantized effective theory where the gauge symmetry is broken, the ``harm'' done by the field redefinition \eqref{CompFields} is arguably still under control. An outline of the argument is as follows. In an effective  QFT, the gauge symmetry breaking anomaly belongs to the cohomology of the  BRST nilpotent operator $s$, whose action on the gauge variables reproduces their infinitesimal gauge transformations, but with infinitesimal gauge parameter replaced by the Fadeev-Popov ghost field $v$ (formally one could write $\gamma=\1+v$ in \eqref{GT}). The nilpotency, $s^2=0$, is secured via the relation $sv=-\tfrac{1}{2}[v,v]=-v^2$ (the commutator is graded w.r.t the form and ghost degrees, $v$ has ghost degree $1$). An anomaly is thus a $n$-form $\A(v, A)$ linear in $v$, which is $s$-closed, $s\A(v, A)=0$, but not $s$-exact, $\A(v,A) \neq sB(A)$. It is determined algebraically through the Stora-Zumino descent equations and receives no contribution from perturbation theory beyond one-loop. Thus, in complete analogy, one could define a ``mock BRST operator'', $\b s$, whose action on the gauge variables produce the infinitesimal version of the composite fields. Formally one could write $u=\1+\upsilon$ in \eqref{CompFields}, where the infinitesimal dressing field satisfies $\b s \upsilon=-\tfrac{1}{2}[\upsilon, \upsilon]=-\upsilon^2$, so that ${\b s}^2=0$. The difference between the effective QFT and its dressed counterpart would be controlled by the ``mock anomaly'' $\b\A(\upsilon, A)$ linear in $\upsilon$, belonging to the $\b s$-cohomology, obtained algebraically via analogues of the Stora-Zumino descent equations and receiving no contribution beyond one-loop from perturbation theory. 


\subsection{Artificial \emph{vs} Substantial Gauge Symmetry} 
\label{Artificial vs substantial gauge symmetry}

Due to the many good properties of gauge theories (notably in relation with renormalizability), over time physicists have devised various ways to implement a gauge symmetry in a theory lacking it \cite{Pitts2008}. 
The Stueckelberg trick is the forefather of these and its generalization seems to  be of some relevance still to contemporary studies \cite{Ruegg-Ruiz}. It is easily illustrated on the historic example of the Proca model (1936) for massive electromagnetism. Proca's Lagrangian is 
\begin{align}
\label{Proca}
L(A)= \tfrac{1}{2}F \w *F+ \mu^2 A\w *A
\end{align}
and describes a massive vector field, so that the theory has no $\U(1)$ gauge symmetry. Now, it is possible to implement such a gauge symmetry by following the suggestion of Stueckelberg (1938) to add a compensating field, the Stueckelberg field $B$ satisfying $B^\gamma=B - \mu \theta$, while the vector field becomes a gauge field transforming as $A^\gamma=A - d\theta$, with $\gamma=e^{i\theta} \in \U(1)$. A minimal Stueckelberg Lagrangian is then,
 \begin{align}
 \label{Stueck}
 L(A, B)=\tfrac{1}{2}F\w*F +   \mu^2 (A-\tfrac{1}{\mu}dB)\w* (A-\tfrac{1}{\mu}dB)
 \end{align}
and is a $\U(1)$-gauge theory. 

In spite of what would be infered from a superficial reading, the Lagrangian \eqref{Proca} and \eqref{Stueck} actually describe the same theory. Indeed the $\U(1)$ gauge symmetry is \emph{artificial}, its presence being compensated by the d.o.f of the field $B$. 
 In the case at hand, the Stueckelberg field is actually an abelian dressing field: $u:=e^{\sfrac{i}{\mu}B}$ so that $u^\gamma=e^{\sfrac{i}{\mu}(B-\mu \theta)}=\gamma\-u$. The associated $\U(1)$-invariant composite fields are then $A^u:=A+iu\-du=A-\tfrac{1}{\mu}dB$,  and its field strength  $F^u=F$. So the Stueckelberg Lagrangian \eqref{Stueck} is rewritten in terms of gauge invariant composite fields as:
\begin{align}
L(A, B)=L(A^u)= \tfrac{1}{2}F^u \w *F^u + \mu^2 A^u\w *A^u,
\end{align}
which is nothing but the Proca Lagrangian, devoid of any gauge symmetry. One may think of the dressing field method as a reciprocal to the Stueckelberg trick: the latter aims at implementing an artificial gauge freedom, the former seeks to erase it to reveal the gauge-invariant content.

The above simple discussion is illustrative of an important point: If one can find in a gauge theory a \emph{local} dressing field, meaning that its value at a spacetime point  depends only on this point and no others,\footnote{This is the notion (2) Field locality, mentioned in the first footnote.} then the invariant composite fields in terms of which the theory can be rewritten are local variables. So, one pays no price in erasing the gauge symmetry, which is then fully dispensable. One therefore proposes the following operational criterion: 
\begin{equation}
\label{C1}
\parbox{.85\textwidth}{
\bf{A local dressing field in a gauge theory signals that  its gauge symmetry is artificial.}} \tag{\text{C1}}
\end{equation}

Gauge theories present a number of conceptual as well as technical challenges. Among those, the fact that the gauge variables have a nondeterministic evolution, and the hindrance gauge symmetry poses a priori to the quantization of the theory. Dirac has pondered long and hard about these difficulties in the context of electromagnetism, an abelian $\U(1)$-gauge theory. A solution he first proposed in a 1955 paper \cite{Dirac55} and then developed in the 1958 fourth edition of his \emph{Principles of Quantum Mechanics} (\cite{Dirac58}, section 80),  was to reformulate the theory with gauge-invariant variables, which would qualify as physical variables.  

In the following we use essentially the notations of \cite{Dirac55} while setting all fundamental constants to $1$. Let $\psi$ be the electron spinor field and $A=(A_0, A_r)$ the electromagnetic gauge potential, subject to the $\U(1)$-gauge transformations $\psi'= e^{iS}\psi$ and $A'= A+dS$.
 Dirac introduces the new variables $\psi^*=e^{iC}\psi$ (Eq [16])  and the associated ``covariant'' derivative $d\psi^*-iA^* \psi^*=e^{iC}(d\psi-iA\psi)$, with $A^*=A+dC$ (Eq [21] and below). The phase factor is defined by $C(x)=\int c_r(x, x') A^r(x')d^3x'$, and in order for the new variables to be gauge invariant $c_r(x,x')$ must satisfy $\tfrac{\d}{\d x'_r} c_r(x,x') =\delta(x-x') $ (Eq [18]). 
 Dirac then notices that the latter equation admits the Coulomb potential as a solution,\footnote{Other solutions differing only by terms dependent on the gauge-invariant Maxwell-Faraday field strength $F$.} so that by proceeding with the quantization of the electromagnetic theory written in terms of his invariant variables, he interprets $\psi^*$ in the following way: 
 \begin{quote}
 ``We can now see that the operator $\psi^*(x)$ is the operator of creation of an electron \emph{together with its Coulomb field}, or possibly the operator of absorption of a positron \emph{together with its Coulomb field}. It is to be contrasted with the operator $\psi(x)$, which gives the creation or absorption of a bare particle. \emph{A theory that works entirely with gauge-invariant operators has its electrons and positrons always accompanied by Coulomb fields around them}, which is very reasonable from the physical point of view.'' 
 \end{quote}
 An appealing conclusion indeed. 

It is not hard to see that Dirac's scheme is an instance of the dressing field method. Indeed under gauge transformation of the gauge potential, the phase factor transforms as
\begin{align*}
C'(x)&=\int c_r(x, x') A^{'r}(x')d^3x'=C(x)+\int c_r(x,x')\tfrac{\d S}{\d x'_{r}}(x')d^3x', \\
       &=C(x)-\int \tfrac{\d}{dx'_r}c_r(x,x')S(x')d^3x'=C(x)-S(x). 
\end{align*}
So  $u=e^{iC}$ transforms under $\gamma=e^{iS} \in \U(1)$ as $u'=\gamma\-u$, and  is therefore an abelian dressing field, which means that $\psi^*$ and $A^*$ in Dirac's equation [16] and [21] are abelian instances of the composite fields $\vphi^u$ and $A^u$ in \eqref{CompFields} above.

Should we then  conclude, on the basis of the criterion \eqref{C1}, that Dirac has revealed the $\U(1)$ gauge symmetry of electromagnetism to be artificial? We must resist that conclusion because here, contrary to what happened in the Stueckelberg model, gauge-invariance wasn't free; it could be achieved only at the price of locality. Indeed the dressing field $u=e^{iC}$ is clearly non-local, so that the gauge-invariant composite fields $\psi^*=\psi^u$ and $A^*=A^u$ in terms of which the theory is rewritten are also non-local variables. 
It appears then that in classical or quantum electrodynamics, there is a trade-off between gauge symmetry and locality: either one works with local gauge variables, or with non-local gauge-invariant ones. 

This conclusion extends to non-abelian Yang-Mills theories. A beautiful articulation is provided by Healey \cite{Healey2009} who argues that the physical content of gauge theories is best represented by the path-ordered trace holonomies of the connection, also known as Wilson loops, which are gauge-invariant non-local variables. The trade-off gauge invariance \emph{vs} locality is indeed a characteristic features of genuine gauge theories (\cite{ Guay2008, Dougherty2017, Nguyenetal2017}), so that one may argue that what is probed, indirectly, by a substantial gauge symmetry is the existence of non-local physical phenomena. In complement to  \eqref{C1}, one therefore proposes the following criterion:
\begin{equation}
\label{C2}
\parbox{.90\textwidth}{
\bf{If only non-local dressing fields exist in a gauge theory, then its gauge symmetry is substantial.}} \tag{\text{C2}}
\end{equation}

 In principle there is an asymmetry: It is easier to verify \eqref{C1} than \eqref{C2}, since for the latter it seems hard to be sure to have exhausted all possibilities in any given case. But an important point should be reminded and stressed at this stage: \emph{The dressing field must be constructed out of the space of primary gauge variables of the theory}. This means that in most relevant cases, the range of options is limited and  for all practical purpose the degree of confidence of having verified \eqref{C2} can be quite high. 
 This also highlights an obvious but crucial fact, namely that the verdict on the nature of the gauge symmetry of a theory depends on its field content.

\medskip

As a manner of illustration, consider
 the Aharonov-Bohm (AB) effect (\cite{AB1959, AB1961}). We recall that  one setup of the effect is a modified double slit experiment involving electrons, $\psi$, where a  solenoid  stands behind the first screen between the two slits. When a current traverses the solenoid, the interference pattern formed by the electrons on the second screen is shifted due to a phase factor depending only on the flux of magnetic field inside the solenoid: $e^{i\int_c A}=e^{i\int_s F}$ ($c$ being a closed path from the source of electron beam through the two slits to a point on the final screen, and enclosing the surface $s$ traversed by the solenoid). Yet, outside the solenoid - the only region accessible to the electrons - the electromagnetic field strength vanishes, $F=0$, and only the electromagnetic potential $A$ is non-zero. So, the latter is the only local variable that is available to maintain a semblance of explanation of the alteration of the behavior of the electrons via a local interaction between two fields, $A$ and $\psi$. 

Of course the gauge non-invariance of $A$ makes it a doubtful candidate as a genuine physical field, as many among  physicists and philosophers alike  have pointed out. Curiously, it is not often stressed that it is also true for the spinor field $\psi$, either seen as a wave function for electrons or as the electron quantum field. Both field variables $A$ and $\psi$ should then be equally faulted for  the difficulty in interpreting the AB effect in terms of local interactions of physical fields.
 Therefore, several authors didn't shy away from concluding that the AB effect forces us to accept, not the physicality of the gauge potential $A$ (which was usually seen as a computational device in classical electromagnetism) as Aharonov and Bohm argued, but rather that there are such things as non-local electromagnetic properties represented by gauge invariant non-local variables.\footnote{As Aharonov and Bohm also argued. To wit, in \cite{AB1961} p.1513 second paragraph:  
 ``The observable physical effects in question must therefore be attributed to the potential integrals themselves. 
Such integrals, being not only gauge invariant, but also Hermitian operators, are perfectly legitimate examples
of quantum-mechanical observables. They represent extended (non-local) properties of the fields [...].'' 
}

But as pointed out by Wallace in \cite{Wallace2014}, in the semi-classical framework where $\psi$ is a complex scalar field, a gauge invariant local interpretation for the AB effect is available. As the reader now acquainted with the dressing field method will recognize, Wallace extracts a \emph{local dressing field} from the polar decomposition $\psi=\rho u$, with $\rho=|\psi|$ and $u= e^{i\theta}$, since indeed $\psi^\gamma=\gamma^{-1}\psi$ implies  $u^\gamma=\gamma^{-1}u$. He then proceeds to show that the theory can be expressed in terms of the gauge invariant local composite fieds $\psi^u=u^{-1}\psi=\rho$ and $A^u=u^{-1}Au-i u^{-1}du=A+d\theta$ (called ``gauge-covariant derivative of $\theta$'' and noted $D\theta$ in \cite{Wallace2014}), with field strength  $F^u=F$, so that the phase factor is $e^{i\int_c A^u}$. Thus, from \eqref{C1} one  concludes that in the scalar-electromagnetic (EM) framework, $\U(1)$ is artificial and the AB effect can be interpreted as resulting from the local interaction of the gauge invariant fields $\psi^u$ and $A^u$.\footnote{It happens that the abelian $\U(1)$ Higgs model can be treated via dressing along the same line, as is shown explicitly in \cite{GaugeInvCompFields}. This is directly relevant to our discussion of the electroweak model in the next section.} 

Notice however that such a ($\U(1)$ adequate) polar decomposition is not available for a Dirac spinor field, so that a local dressing field cannot be thus constructed in the spinorial-EM framework. In this theoretical context, as shown in Dirac's scheme, the only dressing field available is non-local and constructed out of the gauge potential. Hence, via \eqref{C2} one concludes that the  $\U(1)$-gauge symmetry in spinorial-EM is substantial and signals non-local electromagnetic properties, and the AB effect is interpreted as illustrating that fact. 
\bigskip

 The dependence of the nature of a gauge symmetry on the  field content of the theory is further echoed in the theorization of gravitational physics. The gauge structure of General Relativity is most apparent in the tetrad formulation. It is based on the bundle of pseudo-orthonormal frames with the Lorentz group as structure group, $\P(\M, S\!O(1,3))$, endowed with the spin connection $A \in \Lambda^1(\M, \so(1,3))$ whose curvature is the Riemann $2$-form $R(A)=dA+A\w A$. Furthermore, as a frame bundle it comes with a soldering, which is a linear isomorphism between the tangent space at each point $x \in \M$ and Minkoswki space, $\theta : T_x\M \rarrow \RR^{1,3}=\{\RR^4, \eta\}$. It it thus represented by a $1$-form  $\theta \in \Lambda^1(\M, \RR^{1,3})$ written explicitely as $\theta^a={e^a}_\mu dx^\mu$, where ${e^a}_\mu$ is the vierbein, or tetrad field (hence the name of the formulation).
 The Lagrangian of pure gravity GR, $L(\theta, A)$, is invariant under the Lorentz gauge group  $\SO=\left\{ \gamma:M \rarrow S\!O(1,3)\right\}$ acting on the variables as $A^\gamma=\gamma\- A\gamma+\gamma\-d\gamma$ and $\theta^\gamma=\gamma\- \theta$. 
 
 By inspection, it appears clearly that the tetrad field $e={e^a}_\mu$ is a \emph{local} dressing field since it satisfies $e^\gamma=\gamma\- e$. The $\SO$-invariant local composite fields are none other than the linear connection $\Gamma=e\- A e +e\-de$, and its Riemann curvature $R(\Gamma)=e\- R(A)e$.\footnote{As an aside, notice that in line with the general remark in section \ref{The basic mathematical setup of the method} following Eq. \eqref{CompFields}, $\Gamma$ is not in the gauge orbit of $A$ and they should not be seen as related by a mere gauge transformation, as indeed the tetrad $e$ is not an element of the Lorentz gauge group $\SO$.}
  The Lorentz gauge symmetry is then artificial and can be erased at no cost in the theory, whose Lagrangian reduces to the standard Einstein-Hilbert metric formulation: $L(\theta, A)=L_{\text{EH}}(g, \Gamma)$. The metric field, written as $g=e^T\eta e$, indeed appears as a natural $\SO$-invariant field in the Lagrangian. 
   This remains true when gravity couples to radiation (EM), classically described matter (ponderable matter, gas of particles, dust, fluids) or scalar fields, collectively described as $\vphi$. So, according to \eqref{C1}, in general relativistic theories with Lagrangians $L(\theta, A, \vphi)=L(g, \Gamma, \vphi^e)$, the local Lorentz gauge symmetry $\SO$ is artificial and encodes no physics that is not already in the metric formulation (not even torsion or the gravitational AB effect), and doesn't restrict its interpretive resources.
  
  Now, when matter is described by spinor fields, Lorentz gauge symmetry seems no longer dispensable.
 Indeed, the minimal coupling of spinors $\psi$ to gravity is achieved via $A$, more precisely via $\rho(A)$, where $\rho : S\!O(1,3) \rarrow S\!L(2, \CC)$ is a morphism from the Lorentz group to its universal cover: $D\psi=d\psi +\rho(A)\psi$. A Lorentz gauge transformation for the spinor field is then $\psi^\gamma=\rho(\gamma\-) \psi$. There is an induced isomorphism $\b\rho : \RR^{1,3}\rarrow \text{Herm}(2, \CC)$ between Minkoswski space and the space of $2\times2$ hermitian matrices, such that for any $v=v^a \in \RR^{1,3}$ and $S \in S\!O(1,3)$ one has: $\b\rho(Sv)=\rho(S)\b\rho(v)\rho(S)^*$, with $*$ the complex conjugation and transposition. This means in particular that $\b\rho(\gamma\- \theta)=\rho(\gamma\-)\b\rho(\theta)\rho(\gamma\-)^*$, so that the tetrad $\rho(e)$ no longer satisfies the defining property of a dressing field w.r.t $\rho(\SO)$ and cannot be used to dress $\psi$. Thus, in a general relativistic theory with spinors $L(\theta, A, \psi)$, there is no local dressing field fitting for the whole theory, and only non-local dressing fields constructed via the holonomy of the spin connection are likely available. According to \eqref{C2}, the Lorentz gauge symmetry $\SO$ is then substantial.\footnote{This conclusion should be slightly nuanced in light of the following fact. Locally, i.e so long as one works on $U \subset \M$ seen both as a coordinate patch and a trivializing open set for the Lorentz bundle, one can decompose the tetrad as $e=ut$, where $u={u^a}_b \in S\!O(1,3)$ and $t={t^b}_\mu$ has the same d.o.f as the metric field and is such that $g=t^T\eta t$. This decomposition relies on the Schweinler-Wigner orthogonalization procedure, see \cite{GaugeInvCompFields} section 4.3 for details en references. Therefore, one has on the one hand $u^\gamma=\gamma^{-1}u$, so that $u$ is a (minimal) local $\SO$-dressing field. On the other hand, one has $\b\rho(\theta)=\rho(u)\b\rho(t)\rho(u)^*$, and from $\b\rho(\gamma^{-1} \theta)$ one deduces $\rho(u^\gamma)=\rho(\gamma)^{-1}\rho(u)$. So, $u$ is indeed a dressing field suited for the spinor field $\psi$, and one can build - in accordance with \eqref{CompFields} - the  $\SO$-invariant composite spinor field $\psi^u:=\rho(u)^{-1} \psi$, which  couples minimally to gravity via the $\SO$-invariant field $A^u:=u^{-1} A u+u^{-1}du$. Unfortunately, $u$ and by extension the composite fields built from it, depend on the coordinate chart on $\U$ in such a way that they have no determined well-behaved transformation law under coordinate changes. Which makes them ill-defined as global geometrical objects. 

It seems that this construction bears some relation to the attempts - pioneered in the `50s and `60s by DeWitt, Ogievetsky and Polubarinov - to build spinors without introducing tetrads and Lorentz gauge symmetry. In this context $t$ would be called a ``square root'' of the metric. See \cite{Pitts2012} for a nice review with an extensive bibliography. Upon mild restrictions on the admissible coordinate systems, the coordinate transformation law for such spinors is formally attainable, but only in the weak field regime, i.e when $g$ is a small perturbation around $\eta$, and even then the transformation law is metric dependent and highly non-linear. It is thus not yet clear that such a framework is satisfactory in the strong field regime of GR, and that it can be accommodated to QFT in curved spacetime. So, a prudent commitment to the assessment given in the main text seems reasonable.} 

 It is interesting to notice that in both $\U(1)$-EM and $\SO$-gravity, it is the spinorial nature of matter that obstructs the construction of a suitable local dressing field. So that it is the coupled systems ``spinorial matter $+$ EM'' and ``spinorial matter $+$ gravitation'' that display substantial gauge symmetries. 
\bigskip

To wrap up this digression on gravitation, let us consider briefly a final example. In the tentative alternatives to GR known as massive gravity and bi-gravity, general coordinate invariance is broken because beside the usual dynamical metric $g$, a reference metric $f_{ab}$ (often $\eta_{ab}$) is necessary present in the mass term. 
Given any arbitrary coordinate system $\{x^\mu\}$, general covariance is restored via the introduction of scalar fields $\phi^a$ used to define  a vielbein ${{\sf e}^a}_\mu= \tfrac{\d \phi^a}{\d x^\mu}$ that allows to promote the reference metric to a covariant tensor $\b f_{\mu\nu}={\sf e_\mu}^a f_{ab} {\sf e^b}_\nu$ (see \cite{deRham2014} for a review). The  scalar fields $\phi^a$, or the vielbein $\sf e$, are sometimes referred to as  ``clock-fields'' (see \cite{Pitts2008} and references therein) or  ``Stueckelberg fields'', given that they play a role analogous to the Stueckelberg $B$ field used to restore $\U(1)$ in massive EM. It is therefore no surprise that, beyond the mere analogy, the vielbein clock field is also a dressing field in the technical sense. Let $G={G^\mu}_\nu \in GL(4)$ be the Jacobian of a coordinate change, the vielbein transforms as ${\sf e}'={\sf e}G$. So, $u=e\-$ satisfies $u'=G\-u$ and is indeed a local dressing field, which indicates that the general covariance in massive gravity or bi-gravity is artificial and hides the existence of a ``true gauge'': the preferred coordinates $x^\mu=\phi^a$.
\bigskip

The method presented here applies to various situations spanning a significant range of interesting physics. It suggests a criterion that allows to adjudicate the nature of the gauge symmetry of a theory in an almost algorithmic way: Inspect the field content, try to built a local dressing field. If you succeed, \eqref{C1}, then you have shown the gauge symmetry to be artificial and  have found, in the form of the local gauge invariant composite fields \eqref{CompFields}, mathematical objects closer to adequate representatives of the true local physical d.o.f of the theory. If you fail and produce only non-local DF, \eqref{C2}, then you have shown (in all likelihood) the gauge symmetry to be substantial and have located in your theory non-local physical d.o.f  that can be represented by the composite fields \eqref{CompFields}. 

The only remaining non algorithmic part of this scheme is the explicit construction of the dressing field. But given the clear specifications 1) to satisfy the defining gauge transformation property $u^\gamma=\gamma\- u$, and 2) to be constructed out of the space of field variables of the theory, the problem is quite well circumscribed and should not pose much of a challenge in most interesting situations. 

In any outcome, by systematically providing a ``Ockhamized'' version of the theory, the method clarifies its interpretive landscape. 
In particular, as shown by the discussion of the electromagnetic AB effect, given the observations of physical facts, slightly differing formalisms used to predict these facts may sharply differ in the metaphysical interpretations they allow.  

As a case in point, the Glashow-Weinberg-Salam electroweak theory is of particular interest to philosophy of physics.  
 It is therefore apt to analyze it in light of the dressing field method. We do so in the following section.

\section{The Electroweak Theory Without Spontaneous Symmetry Breaking}  
\label{The electroweak theory without spontaneous symmetry breaking}

The opinion that the notion of SSB is pivotal to the success of the  electroweak unification is quite common among physicists. But in the past fifteen years, philosophers of physics started to consider the notion as suspicious. We here argue that their intuitions and conclusions are correct: The SSB interpretation of the electroweak theory is superfluous to its empirical success. Hints at this conclusion were scattered in the gauge field theory literature for years, from the mid-sixties onward, as we will show in the commentary section \ref{There is no SSB in the electroweak model and we long suspected it}. But first, in the following section we prove the main point in sketching  the treatment of the theory via the dressing field method.  Further details and comments can be  found in \cite{Attard_et_al2017}.

\subsection{The electroweak model treated via dressing}
\label{The electroweak model treated via dressing}

The gauge group postulated a priori for the model is $\H=\U(1)\times \SU(2)=\left\{\alpha \times \beta:\M \rarrow U(1)\times S\!U(2)  \right\}$. The space of field is $\Phi=\{A, F, \vphi\}$, where  $A=a+b$ is the gauge potential $1$-form with curvature $F=f_a+g_b$, and $\vphi$ is a $\CC^2$-scalar field.
The latter couples minimally with the gauge potential via the covariant derivative $D\vphi= d\vphi +(g'a +gb)\vphi$, with $g', g$ the coupling constants of $U(1)$ and $S\!U(2)$ respectively. The gauge group  acts as:

\hspace{1.5cm}
\begin{minipage}{.25\linewidth}
\begin{align*}
a^\alpha&=a +\tfrac{1}{g'}\alpha^{-1}d\alpha,\\
a^\beta&=a, 
\end{align*}
\end{minipage}
\begin{minipage}{.25\linewidth}
\begin{align*}
 b^\alpha&=b,   \\
  b^\beta&=\beta^{-1}b\beta + \tfrac{1}{g}\beta\-d\beta,
 \end{align*}
 \end{minipage} 
 \begin{minipage}{.25\linewidth}
 \begin{align*}
\text{and} \quad \vphi^\alpha &=\alpha^{-1}\vphi,  \notag\\
 \text{and} \quad\vphi^\beta&=\beta^{-1}\vphi.
\end{align*}
\end{minipage}
\bigskip

\noindent The $\H$-invariant Lagrangian form of the theory is, 
\begin{align}
\label{EW-Lagrangian}
L(a, b, \vphi)&=\tfrac{1}{2}\Tr(F \w * F) + \langle D\vphi, \ *D\vphi\rangle - U(\vphi) \vol, \notag\\
			&=\tfrac{1}{2}\Tr(f_a\wedge *f_a) +\tfrac{1}{2}\Tr(g_b\wedge *g_b) + \langle D\vphi, \ *D\vphi\rangle -\left( \mu^2 \langle\vphi, \vphi\rangle +\lambda \langle\vphi, \vphi\rangle^2\right) \vol,
\end{align}
where $\mu, \lambda \in \RR$ and $\vol$ is the volume form on spacetime $\M$. 
As it stands, nor $a$ nor $b$ can be massive, and indeed $L$ contains no mass terms for them. While at least one massless field is expected in order to carry the electromagnetic interaction,  the weak interaction is short range so its associated fields must be massive. Hence the necessity to reduce the $\SU(2)$ gauge symmetry  in order to allow for  mass terms for the weak fields. 

As the usual narrative goes, this is achieved via SSB: If $\mu^2<0$, the electroweak vacuum given by $U(\vphi)=0$ seems degenerate as it appears to be an $\SU(2)$-orbit of non-vanishing vacuum expectation values for $\vphi$. When the latter settles randomly - spontaneously - on one of them, this breaks $\SU(2)$ and generates mass terms for the weak fields with which it couples. Oddly, in order to exhibit the physical modes of the theory it is claimed that a convenient choice of gauge is necessary, the so-called \emph{unitary gauge} (see e.g \cite{Becchi-Ridolfi}). But how come we are allowed to use a gauge freedom if it is supposedly broken? 
\medskip

We suggest that a better approach and a more satisfactory interpretation is provided by the dressing field method. 
Indeed it is not hard to find a dressing field in the electroweak model.
 Considering the polar decomposition in $\CC^2$ of the scalar field $\vphi=u \eta$ with
\begin{align}
\label{decomp-phi}
    u\in S\!U(2) \quad \text{ and }\quad \eta:=  \vect{0 \\ ||\vphi||} \in \RR^+ \subset \CC^2, \qquad \text{one has} \quad \vphi^\beta  \quad    \Rightarrow  \quad   u^\beta=\beta\- u.
\end{align}
Thus,   $u$ is a  $\SU(2)$-dressing field that can be used to construct the $\SU(2)$-invariant composite fields:
\begin{align}
 A^u&=u\-Au+\tfrac{1}{g}u\-du= a+  (u\-bu +\tfrac{1}{g}u\-du)= a + B,     \notag \\
 F^u&=u\- F u=f_a+u\-g_bu =f_a+G,\qquad \textrm{with $G=dB+gB^2$,}     \notag\\[1mm]
 \vphi^u&=u\-\vphi=\eta, \qquad \text{and} \qquad  D^u\eta=u\-D\vphi=d\eta + (g'a + gB)\eta.
\end{align}
Since $u$ is \emph{local}, so are the above composite fields. Therefore, by  virtue of criterion \eqref{C1} we conclude that the $\SU(2)$-gauge symmetry of the model is \emph{artificial}, so that the theory defined by the electroweak Lagrangian \eqref{EW-Lagrangian} is actually a $\U(1)$-gauge theory, described in terms of local $\SU(2)$-invariant variables:
\begin{align}
\label{EW-Lagrangian2}
L(a, B, \eta)=\tfrac{1}{2}\Tr( F^u \w *F^u) +  \langle D^u\eta, \ *D^u\eta \rangle -\left( \mu^2\eta^2 +\lambda \eta^4  \right) \vol.
\end{align}

We already reached our main conclusion:  Since the $\SU(2)$-gauge symmetry is artificial, the interpretation of the model in terms of SSB is superfluous, and indeed impossible when expressed in the form \eqref{EW-Lagrangian2}. We could then stop here. But at this point it is not clear that as it stands our analysis reproduces all the phenomenology usually obtained via the standard interpretation. In what follows we show that it is indeed so: It is done simply by proceeding to the natural step of analyzing the residual and substantial $\U(1)$ gauge symmetry of the model, which is very easily done  from the viewpoint of the dressing field method.

\paragraph{Residual $\U(1)$ symmetry}

By its very definition $\eta^\beta=\eta^\alpha=\eta$, so it is already a fully $\H$-gauge invariant scalar field which  then qualifies as an observable. 
As a rule, the  $\U(1)$-residual gauge transformation of the  $\SU(2)$-invariant composite fields depends on the $\U(1)$-gauge transformation of the dressing field $u$. One finds that 
\begin{align*}
\vphi^\alpha \quad \Rightarrow \quad u^\alpha= u\t\alpha, \qquad \text{where} \qquad  \t\alpha= \begin{pmatrix} \alpha & 0  \\ 0 & \alpha\- \end{pmatrix}.
\end{align*}
Therefore, 
$B^\alpha=(b^\alpha)^{u^\alpha}=b^{u\t\alpha}=\t\alpha\-u\- b u\t\alpha +\tfrac{1}{g}\t\alpha\-(u\-du)\t\alpha +\tfrac{1}{g}\t\alpha\-d\t\alpha$ $
=\t\alpha\- B \t\alpha + \tfrac{1}{g}\t\alpha\-d\t\alpha.$ 
Given the decomposition $B=B_a \sigma^a$, where $\sigma^a$ are the hermitian Pauli matrices and $B_a \in i\RR$, we have explicitly
\begin{align*}
B=\begin{pmatrix} B_3 & B_1-iB_2 \\ B_1+iB_2 & -B_3  \end{pmatrix}=:\begin{pmatrix} B_3 & W^- \\ W^+ & -B_3  \end{pmatrix}, \quad \textrm{and}  \qquad B^\alpha=\begin{pmatrix}  B_3 +\frac{1}{g}\alpha\-d\alpha & \alpha^{-2}W^- \\[5pt] \alpha^2W^+ & -B_3 -\frac{1}{g}\alpha\-d\alpha \end{pmatrix}.
\end{align*}
The fields $W^\pm$ transform tensorially, it is then possible  for these two fields to be massive. $B_3$ transforms as a $\U(1)$-gauge potential, but with a different coupling constant, making it another massless field together with the initial $\U(1)$-gauge potential $a$. 
But consider the $\U(1)$-transformation of the $\SU(2)$-\textit{invariant} covariant derivative:
\begin{align*}
D^u\eta= d\eta+(g'a + gB) \eta =\vect{gW^-\eta \\ d\eta - gB_3\eta +g'a\eta},\quad \textrm{ and}\quad (D^u\eta)^\alpha=\vect{g\alpha^{-2}W^-\eta \\ d\eta - gB_3\eta +g'a\eta}.
\end{align*}
We see that a $\U(1)$-invariant combination of $a$ and $B_3$ appears. So, considering $(a, B_3)$ as a doublet in $
\CC^2$, one is invited to perform the natural change of variables
\begin{align*}
\vect{A \\ Z^0} =\begin{pmatrix} \cos\theta_W &  \sin\theta_W \\ -\sin\theta_W & \cos\theta_W \end{pmatrix} \vect{a \\ B_3} = \vect{\cos\theta_W a + \sin\theta_W B_3\\  \cos\theta_W B_3 - \sin\theta_W a},
\end{align*}
where  $\cos \theta_W = \sfrac{g}{\sqrt{g^2+g'^2}}$ and $\sin\theta_W=\sfrac{g'}{\sqrt{g^2+g'^2}}$ ($\theta_W$ is known as the Weinberg, or weak mixing, angle).
By construction, the $1$-form $Z^0$  is fully $\H$-gauge invariant, thus observable and potentially massive.  
Now, still by construction,  $A^\beta=A$ and  $A^\alpha=A + \tfrac{1}{e} \alpha\-d\alpha$ with coupling constant $e=\sfrac{gg'}{\sqrt{g^2+g'^2}}$.
So, $A$  transforms as a $\U(1)$-gauge potential, it can thus be interpreted as the massless mediator of the electromagnetic interaction whose coupling constant  $e$ is the elementary electric charge.

The electroweak theory \eqref{EW-Lagrangian2} is then expressed in terms of the gauge invariant fields $\eta, Z^0$ and of the $\U(1)$-gauge fields $W^\pm, A$:
\begin{align*}
L(A, W^\pm, Z^0, \eta)= \ \tfrac{1}{2}\Tr(F ^u\w * F^u) + d\eta\wedge*d\eta  -  g^2\eta^2\  W^+\wedge*W^-    -  (g^2+g'^2)\eta^2\  Z^0\wedge*Z^0 	 -\left( \mu^2\eta^2 +\lambda \eta^4  \right) \vol.
\end{align*}

The next natural step is to expand the $\RR^+$-valued scalar field $\eta$ around its \emph{unique} groundstate $\eta_0$,\footnote{According to  \cite{Westenholz} the very meaning of the terminology ``\textit{spontaneous} symmetry breaking'' lies in the fact that the manifold of vacua is not reduced to a point.} given by $U(\eta_0)=0$, as $\eta=\eta_0+H$ where $H$ is the gauge invariant Higgs field. Mass terms for  $Z^0, W^\pm$ and $H$ depending on $\eta_0$  appear from the couplings of the electroweak fields with $\eta$ and from the latter's self interaction.\footnote{Since $A$ does not couple to $\eta$ directly it is massless. The two photons decay channel of the scalar boson involves intermediary leptons, not treated here but whose inclusion in this scheme is straightforward.}  The theory has two qualitatively distinct phases. 
In the phase  where $\mu^2>0$,  $\eta_0$ vanishes and  so  do all masses. But in the phase where $\mu^2<0$, the groundstate is non-vanishing: $\eta_0=\sqrt{\sfrac{-\mu^2}{2\lambda}}$.  The masses of the fields $Z^0, W^\pm$  and $H$ are then
$m_{Z^0}=\eta_0\sqrt{(g^2+g'^2)}$,  $m_{W^\pm}=\eta_0g$  and $m_H=\eta_0\sqrt{2\lambda}$.  
All physical predictions of the electroweak theory are indeed preserved in our treatment: masses are gained through a phase transition of the unique electroweak vacuum. 
 
In conclusion, notice that this approach to the electroweak model offers a satisfactory reconciliation with  Elitzur's theorem (\cite{Elitzur1975}), stating that in lattice gauge theory a gauge symmetry cannot be spontaneously broken.

 \subsection{There Is No SSB In The Electroweak Model And We Long Suspected It}
\label{There is no SSB in the electroweak model and we long suspected it}

 It turns out that several authors were close to formulating such a gauge invariant account of the electroweak theory. Even before the theory was proposed, in 1965 - barely a year after his celebrated paper - Higgs hinted at a gauge invariant formulation of the mechanism that ended-up bearing his name by working on an abelian toy model, see section IV in \cite{Higgs66}.  In 1966, two years after his own celebrated contribution with Guralnik and Hagen, Kibble  \cite{Kibble67} suggested a similar analysis but working on both abelian and non-abelian models. Just before the conclusion of his paper he writes:
 \begin{quote}
  ``We note certain characteristic features of our model. It is perfectly possible to describe it without ever introducing the notion of symmetry breaking, merely by writing down the Lagrangian (66) [written with gauge invariant variables]. Indeed if the physical world were really described by this model, it is (66) rather than (64) [i.e, the Lagrangian written in terms of gauge variables] to which we should be led by experiment.''
 \end{quote}
 
 With insight it is clear that both Higgs and Kibble were using instances of the dressing field method: see equation (22) and below  (dressing and composite fields) in \cite{Higgs66}
   as well as equations (9)-(59) (dressing fields) and (16)-(61) (composite fields) in \cite{Kibble67}. 
   
But then the Glashow-Weinberg-Salam model was proposed the next year, using the BEHGHK\footnote{Brout-Englert-Higgs-Guralnik-Hagen-Kibble, to honor all contributors.} mechanism with its original  interpretation in terms of SSB. So these important insights from Higgs and Kibble were eclipsed, and the view of the SSB a real physical phenomenon which had happened in the early universe gained currency. In  a panel discussion during a large conference on the foundation of quantum field theory gathering physicists as well as historians and philosophers of physics in Boston in 1996, the following exchange took place \cite{Cao1999}:

\begin{quote}
%
%
%
%
%
%
%

Nick Huggett: [...] what is the mechanism, the dynamics for spontaneous symmetry breaking supposed to be? [...] My worry is there's supposed to be a transition from an unbroken symmetry to the [current] state [...] isn't this a dynamic evolution, something that happens in the history of the universe? 
[...]

Sidney Coleman: Yes, typically at high temperature the density matrix has a symmetry which then disappears as the temperature gets lower. But its also true for ordinary material objects. [...] The difference between the vacuum and every other quantum mechanical system is that it's bigger. And that's from this viewpoint the only difference. If you understand what happens to a ferromagnet when you heat it up above the Curie temperature, you're a long way towards understanding one of the possible ways it can happen to the vacuum state. 
\end{quote}

Yet, the treatment of the electroweak model through the bundle reduction theorem - see e.g \cite{Trautman, Westenholz, Sternberg} - already cast some doubts on the interpretation of the SSB as a dynamical phenomenon. 
Indeed, thus formulated it appears that the model can naturally be rewritten on a $U(1)$-subbundle of the initial $U(1)\times S\!U(2)$-bundle.


As far as I know the first to give a fully $\SU(2)$-gauge invariant formulation of  the electroweak theory were Fröhlich, Morchio and Strocchi in 1981 (\cite{Frohlich-Morchio-Strocchi81}). Their account it actually fully equivalent to ours, but much less synthetic and systematic: They are working on individual scalar components of all the fields involved! See their equations (6.1) describing the composites fields (including dressed electron and neutrino). Subsequently, and especially in the last 10 years, several researchers independently
rediscovered the gauge invariant description formulated essentially as in the above treatment, but without the conceptual clarity given by the dressing field method, as often the dressing field was mistaken for an element of the gauge group   (\cite{McMullan-Lavelle95}, see in particular equations (6)-(7) and comment in between) or  the interpretive shift was not fully embraced  (\cite{Chernodub2008, Faddeev2009}). Interestingly, the textbook by Rubakov gives essentially the dressing treatment of the abelian Higgs model, but sticks to the usual treatment of the electroweak model using the unitary gauge, see \cite{Rubakov1999} chapter 6. Some improvement in conceptual clarity is found in \cite{Ilderton-Lavelle-McMullan2010}. But it is  Masson and Wallet \cite{Masson-Wallet} who  first  really appreciated the interpretive shift that comes with the invariant formulation, and as a matter of fact their paper was a precursor to the development of the dressing field method. Unfortunately it never get published.
 \medskip
 
 In parallel, in the last fifteen years, philosophers of science have questioned the orthodoxy of the SSB in the electroweak model.
 Earman first raised the issue in striking terms: 
  \begin{quote}
``But what exactly is accomplished [in the BEHGHK mechanism]  is hidden behind the veil of gauge
redundancy. The popular presentations use the slogan that the vector field has
acquired its mass by `eating' the Higgs field. [...]
The popular slogan can be counterbalanced by the cautionary
slogan that neither mass nor any other genuine attribute can be gained by eating
descriptive fluff.
None of this need be any concern for practicing physicists who know when they
have been presented with a fruitful idea and are concerned with putting the idea to
work. But it is a dereliction of duty for philosophers to repeat the physicists' slogans
rather than asking what is the content of the reality that lies behind the veil of gauge.''  
\vspace{-2mm}
 \flushright{\cite{Earman2004a} pp189-190.}
  \end{quote}
 Shortly after, in \cite{Earman2004b}, he reiterated that ``a genuine property like
mass cannot be gained by eating descriptive fluff, which is just what gauge
is. Philosophers of science should be asking the Nozick question: What
is the objective (i.e., gauge invariant) structure of the world corresponding
to the gauge theory presented in the Higgs mechanism?''. 
  Emphasizing Dirac's constrained Hamiltonian formalism as a systematic way to extract the gauge invariant quantities of a gauge system he asks: 
\begin{quote}
``What is the upshot of applying this reduction procedure to the Higgs model and
then quantizing the resulting unconstrained Hamiltonian system? In particular, what
is the fate of spontaneous symmetry breaking? To my knowledge the application has
not been carried out. [...] 

While there are too many what-ifs in this exercise to allow any firm conclusions to
be drawn, it does suffice to plant the suspicion that when the veil of gauge is lifted,
what is revealed is that the Higgs mechanism has worked its magic of suppressing zero
mass modes and giving particles their masses by quashing spontaneous symmetry
breaking. However, confirming the suspicion or putting it to rest require detailed
calculations, not philosophizing.'' 
\vspace{-2mm}
 \flushright{\cite{Earman2004a} pp190-191.}
\end{quote}

Since then, several authors have raised to the challenge,  noticing or rediscovering for themselves the invariant formulation (\cite{Smeenk2006, Lyre2008, Struyve2011, Friederich2013, Friederich2014}) and all essentially concluded that gauge SSB is indeed a dispensable notion. 
Since they rely to some significant degree on the dressing field method, the considerations presented in section \ref{The electroweak model treated via dressing} fully confirm their conclusion in synthesizing the core technical argument, and vindicates Earman's suspicion. Furthermore, it clearly places the specific question of gauge SSB in the electroweak model  
within the broader problem of distinguishing substantial from artificial gauge symmetries.

\section{Closing statement, open questions}
\label{Closing statement, open questions}

We have seen that if it is usually recognized that an important demarcation criterion between artificial and substantial symmetries is that the former can be erased without forsaking the locality of a theory while this is not so for the latter, in practice the distinction is not so readily recognized.
The dressing field method is a general tool that allows to systematically implement that criterion. If a gauge theory contains a suitable local dressing field, it can be rewritten in terms of local gauge invariant composite fields. Nothing is then lost in erasing the gauge symmetry, so one can argue that it was artificial, stemming from an uneconomical - a ``non-Ockhamized'' - choice of variables. If on the contrary a gauge  theory contains only  non-local dressing fields, then its gauge symmetry is erased at the price of a rewriting  in terms of non-local gauge invariant variables. One then conclude that the gauge symmetry of the theory is substantial, and may signal the existence of non-local physical phenomena (as the analysis of the AB effect in spinorial EM exemplifies).
Furthermore, the dressing field method highlights the - subtle or obvious - fact that the verdict on the nature of a gauge symmetry, and therefore the available interpretations of the theory, crucially depend on the field content of the theory. Our discussion of the AB effect illustrates that point.
In particular, the treatment of the electroweak model via the dressing field method shows that the $\SU(2)$ gauge symmetry is artificial, canceling the need for the  notion of gauge SSB. Only the residual $\U(1)$ gauge symmetry is substantial. 
Provocatively, one could say that the substantial gauge group of the Standard Model of particle physics is therefore not $\U(1)\times\SU(2)\times\SU(3)$, but merely $\U(1)\times\SU(3)$.
\medskip

It is the job of both mathematical physicists and philosophers of physics to prune a theory from any superfluous notion that pertains to the context of discovery so as to reveal its core conceptual and technical structure, and to clear the horizon of its context of justification. 
Here we conclude that the notion of gauge SSB pertains to the context of discovery of the electroweak unification: It has historical interest and has been a valuable heuristic guide to the correct theory, but  it cannot belong to the context of justification. 

A puzzling facts  remains: How are we to understand that the artificial $\SU(2)$ formulation of the model - such as suggested by the gauge principle - is structurally much more  simple  than the substantial $\U(1)$ - and phenomenologically clearer - formulation? Let us continue the quotation from Kibble \cite{Kibble67}:
\begin{quote}
``Indeed if the physical world were really described by this model, it is (66) [the Lagrangian written with gauge-invariant variables] rather than (64) [the Lagrangian exhibiting a gauge symmetry] to which we should be led by
experiment. The only advantage of (64) is that it is
easier to understand the appearance of an exact symmetry than of an approximate one. Experimentally, we
would discover the existence of a set of four vector
bosons with different masses but whose interactions
exhibited a remarkable degree of symmetry. We would
also discover a pair of scalar particles forming an
apparently incomplete multiplet under the group describing
this symmetry. In such circumstances it would
surely be regarded as a considerable advance if we could
recast the theory into a form described by the symmetric
Lagrangian (64).'' 
\end{quote}
But given that the gauge symmetry is in this circumstance artificial, it must me clarified in what respect it is an advance.  Furthermore, if $\SU(2)$ is artificial and as such should not tell us anything important, it is a remarkable feat that the model guessed from it eventually had such predictive power.\footnote{The model was proposed in 1967 and predicted the neutral weak current, a specific relation between the masses of the $W^\pm$ and  the $Z^0$ as well as the existence of a scalar boson. The neutral weak current was discovered in 1973 in the Gargamelle experiment at CERN. The $W^\pm$ bosons were discovered in January 1983 and the $Z^0$ boson in May 1983, also at CERN. Finally the discovery of the scalar boson at the LHC was announced in July 2012.}
Are we to believe that the distinction artificial \emph{vs} substantial gauge symmetry does not capture  all important theoretical differences and must be reconsidered? 

I find this unlikely. My guess is that it remains to determine what constitutes the proper context of justification for the electroweak theory.  The gauge principle associated with the substantial $\U(1)$-symmetry is clearly insufficient. And if a phenomenological a posteriori reconstruction is possible, it does not illuminates the key ideas or principles that might explain the structure of the theory. Actually the question stands: Is there a  principle that would make the theory something other than a raw fact? Renormalizability of the quantum theory may come to mind as a powerful constraining factor, but is it to be elevated to such a high position in the explicative hierarchy? Effective field theory physicists would disagree. 
 Interestingly, it can be shown that the requirement that vector fields  interacting with spinor and scalar fields have spin 1 leads naturally to the group structure of interactions characteristic of Yang-Mills theories (\cite{OP1963, OP1964, OP1966}). For massless vector fields this requirement is equivalent to imposing a gauge symmetry. One could then consider that it is the notion that weak fields have spin 1 that explains the $\SU(2)$-gauge structure of the electroweak model, rather than postulating the gauge structure as explaining the weak interactions in terms of spin 1 fields. 
This would partly alleviate the puzzlement on why an artificial $\SU(2)$ symmetry turned out to be a good heuristic guide: It flows ``accidentally'' from a reasonable physical ``spin 1 principle''. 
But then, this principle produces Lagrangians in the ``undressed'' form  displaying both artificial and substantial gauge symmetries, and is therefore blind to the difference. It is still surprising that a principle producing directly the empirically clear $\U(1)$-substantial form of the theory is elusive.

It is admitted that the Standard Model should be a low energy limit of a more fundamental theory. The governing principle we search for may  be part of the new framework within which this fundamental theory is expressed. Could it be a new geometric framework, such as non-commutative geometry or transitive Lie algebroids?  Could it be a firmer mathematical foundation for quantum field theory, such as the algebraic formulation or category theory? Reversing the logic, it may be that pondering on what explains the form of the electroweak unification could provide hints on this as yet unidentified framework and on what lies beyond the Standard Model. 
These questions can be genuinely explored only if the orthodoxy of SSB, a context of justification turned into a common wisdom, is challenged. Philosopher of physics have spearheaded that challenge in the past fifteen years. It is to be hoped that the community of physicists catches up quickly.

%

{
\normalsize 
 \bibliography{Biblio}
}

\end{document}